\documentclass[sigconf]{acmart}

\setcopyright{acmcopyright}
\copyrightyear{2019}
\acmYear{2019}
\acmDOI{10.1145/1122445.1122456}

\acmConference[arXiv]{arXiv Preprint}{ December 2019}{ Rice University }
\acmBooktitle{arXiv Preprint}
\acmPrice{15.00}
\acmISBN{ }
\acmDOI{ }
\setcopyright{none}
\renewcommand\footnotetextcopyrightpermission[1]{}
\usepackage{hyperref}
\usepackage{url}
\usepackage{algorithmic}
\usepackage{algorithm}
\usepackage{makecell}
\usepackage{graphicx}
\usepackage{amsmath}
\usepackage{amsthm}
\usepackage{amsfonts}
\newtheorem{definition}{Definition}
\newtheorem{theorem}{Theorem}
\newtheorem{corollary}{Corollary}
\newcommand\KDE{\mathrm{KDE}}
\DeclareMathOperator{\E}{\mathbb{E}}
\DeclareMathOperator{\R}{\mathbb{R}}
\DeclareMathOperator{\var}{\mathrm{var}}

\begin{document}

\title{Sub-linear RACE Sketches for Approximate Kernel Density Estimation on Streaming Data}


\author{Benjamin Coleman}
\affiliation{%
  \institution{Department of Electrical and Computer Engineering \\ Rice University}
  \streetaddress{6100 Main Street}
  \city{Houston}
  \country{Texas}
  \postcode{77005}}
\email{brc7@rice.edu}

\author{Anshumali Shrivastava}
\affiliation{%
  \institution{Department of Computer Science \\ Rice University}
  \streetaddress{6100 Main Street}
  \city{Houston}
  \country{Texas}
  \postcode{77005}}
\email{anshumali@rice.edu}



\begin{abstract}
Kernel density estimation is a simple and effective method that lies at the heart of many important machine learning applications. Unfortunately, kernel methods scale poorly for large, high dimensional datasets. Approximate kernel density estimation has a prohibitively high memory and computation cost, especially in the streaming setting. Recent sampling algorithms for high dimensional densities can reduce the computation cost but cannot operate online, while streaming algorithms cannot handle high dimensional datasets due to the curse of dimensionality. We propose RACE, an efficient sketching algorithm for kernel density estimation on high-dimensional streaming data. RACE compresses a set of $N$ high dimensional vectors into a small array of integer counters. This array is sufficient to estimate the kernel density for a large class of kernels. Our sketch is practical to implement and comes with strong theoretical guarantees. We evaluate our method on real-world high-dimensional datasets and show that our sketch achieves 10x better compression compared to competing methods.
\end{abstract}

\ccsdesc[500]{Theory of computation~Sketching and sampling}



\keywords{kernel density estimation, sketching, streaming data, locality sensitive hashing, compression}

\maketitle

\section{Introduction}
Kernel density estimation (KDE) is a well-studied classical nonparametric approach to directly estimate the probability density function from a given set of data points. The method is a powerful modeling tool at the core of many data analysis tasks, including regression~\cite{chen2018modal}, anomaly detection~\cite{schubert2014generalized}, mode finding~\cite{carreira2000mode}, visualization~\cite{lampe2011interactive}, classification~\cite{john1995estimating}, clustering~\cite{hinneburg2007denclue,kriegel2011density} and more. Kernel density estimates enable such a diverse range of applications because they are a simple way to represent the data distribution. 

Given a dataset $\mathcal{D} = \{x_1, x_2, ... x_N\} \subset \R^d$, a positive semi-definite kernel $k(\cdot,\cdot)$, and a query $q$, the kernel density estimate (KDE) at $q$ is defined as $\KDE(q) = \frac{1}{N} \sum_{x\in \mathcal{D}} k(x,q)$. Exact computation of the KDE requires $O(N)$ evaluations of the kernel function and $O(Nd)$ memory, which is prohibitively expensive in practice. Therefore, it is common to solve a relaxed version of the problem where the objective is to efficiently provide a $1\pm \epsilon$ multiplicative approximation to the KDE. Such an approximation is sufficient for the vast majority of applications. 

In this paper, our focus is KDE in memory constrained settings where it is prohibitive to store the complete data in any form. This restriction applies to resource-constrained settings such as the Internet-of-Things (IoT), communication-constrained settings such as computer networks, and distributed database settings in genomics and web-scale machine learning. 

\paragraph{Formal Description of the Compressed KDE Framework} Given an $N$-point dataset $\mathcal{D} = \{x_1, x_2, ... x_N\} \subset \R^d$ of vectors observed in a one-pass sequence, the task is to build a small data structure $\mathcal{S}$ that can efficiently estimate the KDE for any dynamically generated query $q$. The algorithm must operate in the strict one-pass streaming setting~\cite{fiat1998online}. $\mathcal{S}$ must support incremental addition and deletion of vectors, and the size of $\mathcal{S}$, which determines the storage and communication cost, should scale $< O(N)$. Since our goal is to explicitly estimate the pairwise kernel summation, approaches which estimate high-dimensional distributions using other, non-kernel methods ~\cite{majdara2019online,lu2013multivariate} are not directly applicable. We restrict our attention to pairwise kernel summations because they are a common and useful subroutine used by a variety of machine learning algorithms. 

In this paper we present an algorithm that solves the compressed KDE problem by representing $\mathcal{D}$ as a tiny sketch. The sketch is a collection of integer counters. Surprisingly, we only store the integer counters and do not need to store any of the $d$-dimensional data vectors in any form. The algorithm is easy to implement, computationally efficient and trivially parallel. Furthermore, the sketches presented in this paper are also \textit{linear} and \textit{mergeable} according to the definitions given in~\cite{agarwal2013mergeable}. That is, given two datasets $\mathcal{D}_1$ and $\mathcal{D}_2$, the sketch of the combined dataset $\mathcal{D}_1 \cup \mathcal{D}_2$ is $\mathcal{S}_1 + \mathcal{S}_2$ with the same error guarantee enjoyed by the original sketches. These unique properties make our method well-suited to kernel density estimation in distributed settings.


\subsection{Applications}
Here, we describe two applications that can directly benefit from our compressed KDE framework. 

\paragraph{Online Dataset Summaries}

Given an efficient KDE oracle, we can perform clustering, modal regression, classification and a mature set of other machine learning kernel methods. Therefore, any sketch that preserves the KDE is clearly an informative representation of a dataset. We propose that RACE sketches can be used to represent large datasets in limited space while preserving the most salient properties of the data. It is not difficult to find applications for such a kernel summary. For instance, web-scale user modeling applications often collect high dimensional feature vectors that represent user behavior or activity. 
In bioinformatics, the state-of-the-art PromenthION sequencing machine streams terabytes of genetic data every day. IoT systems can generate streams of thousands of distributed signals.

All of these applications involve problems where pairwise kernel summations are bound by computation and communication costs. Applications at this scale require efficient data summaries for kernel density that work in the streaming setting, since it is impractical to store the complete data in any form~\cite{qin2019scalable}. Our method provides a computationally efficient approach for sketching the KDE on web-scale streaming data. 


\paragraph{Compressed Classification in Networks} Online classification is a common task for sensor networks in IoT systems and internet server networks. Consider a system where a sensor produces a stream of raw data that must be relayed back to a processing center. Alternatively, consider a distributed server network where each server needs to communicate information about its workload to other nodes. The server infrastructure at AT\&T is a real-world example where kernel density sketches are used to analyze server load in communication-constrainted settings ~\cite{procopiuc2005density}. Computation and communication are major bottlenecks for performing online classification on network and sensor data. 

Since KDE is a natural way to implement likelihood ratio classifiers, RACE sketches can provide an efficient online classification algorithm. RACE summaries are linear and mergeable, allowing them to work in the distributed network setting described by~\cite{procopiuc2005density}. 
Furthermore, our sketch is hardware and FPGA-friendly as it consists entirely of fixed-size integer arrays. We anticipate that RACE sketches will enable fast and memory-efficient maximum likelihood classification on streaming data. 



\subsection{Related Work}
Advances in KDE have been historically motivated by runtime due to the prohibitive $O(N)$ query evaluation cost. As a result, time-efficient KDE on low-dimensional data is a very well-studied problem, with many practical solutions based on the fast multipole and dual tree methods. Unfortunately, most KDE methods require $O(N)$ memory and are not applicable to our situation. The memory footprint of KDE has recently begun to receive attention~\cite{siminelakis2019rehashing}, but primarily in the context of sketch-based methods that improve the runtime. While low-memory KDE sketches are not particularly well-studied, several existing approaches are good candidates for sketching the KDE on streaming data.

\begin{table*}[t]
\begin{center}
\begin{tabular}{ l | l | l }
\hline
Method & Memory & Comments\\
\hline
\makecell[l]{Kernel Merging ~\cite{zhou2003m,heinz2008cluster,cao2012somke}} & $O(d M)$ & \makecell[l]{$M$ is the number of merged kernels, determined \\ empirically. Can be constructed and queried online. }\\
\makecell[l]{Random Sampling~\cite{muandet2017kernel} } & $O\left( \frac{1}{\KDE}\frac{d}{\epsilon^2} \log \frac{1}{\delta}\right)$ & \makecell[l]{Widely used in practice. Can be constructed online \\ using reservoir sampling. }\\
\makecell[l]{Hash-Based Sketch~\cite{siminelakis2019rehashing}} & $O\left(\frac{1}{\KDE}\frac{d}{\epsilon^2}\log \frac{1}{\delta} \right)$ & \makecell[l]{LSH-based method with 2-pass construction. \\ Constant factors are better than random sampling.}\\
\makecell[l]{Kernel Herding~\cite{chen2012super}} & $O\left(\frac{1}{\KDE}\frac{d}{\epsilon}\right)$ & \makecell[l]{Requires computing the exact $\KDE$ for all points in $\mathcal{D}$. \\ Prohibitive $O(|\mathcal{D}|^2)$ sketching cost. }\\
\makecell[l]{Sparse Kernel Approximation~\cite{cortes2016sparse}} & $O(d(1 - \frac{\epsilon}{\alpha}\KDE)|\mathcal{D}|)$ & \makecell[l]{$\alpha$ is a dataset-dependent constant $\leq 1$. Requires a \\ matrix inversion that scales poorly with sketch size }\\
\makecell[l]{Coresets~\cite{phillips2018near}} & $O\left( \frac{1}{\KDE}\frac{\sqrt{d^3}}{\epsilon}\sqrt{\log{\frac{1}{\epsilon\KDE}}}\right)$ & \makecell[l]{ $O(|\mathcal{D}|^2)$ sketching cost. Alternative methods have \\ different dependencies on $\epsilon\KDE$ (See \cite{phillips2018near})}\\
\makecell[l]{RACE} & $O\left( \frac{1}{\epsilon^2}\frac{1}{\KDE^2} \log \frac{1}{\delta}\right)$ & \makecell[l]{This work. Results apply for all LSH kernels with a tighter \\ bound for a subset of kernels.}\\
 \hline
\end{tabular}
\end{center}
\caption{Summary of related work. In this table we consider the problem of estimating $\KDE(q)$ for a dataset $\mathcal{D}\in \R^d$ with multiplicative error $\epsilon$. The algorithm is allowed to fail with probability $\delta$. The guarantees for coresets, sparse kernel approximation and kernel herding do not fail with probability $\delta$ as these algorithms are deterministic. We assume the kernel induces a reproducing kernel Hilbert space (RKHS) for \cite{chen2012super} and \cite{cortes2016sparse}. RACE is the only method which does not scale with the per-sample storage cost $d$.}
\label{table:relatedwork}
\end{table*}

\paragraph{Merging} Suppose we cluster the dataset to obtain a set of cluster centroids. A reasonable KDE approximation is to report a weighted KDE over the centroids, where each weight is determined by the size of each cluster. This approach reduces the computation and memory needed to report the KDE by collapsing data points to the nearest centroid. While this approach requires multiple passes through the data, most online KDE methods rely on the same intuition. Existing algorithms for online KDE evaluation are based on \textit{kernel merging}. These methods begin by allocating a buffer of $M$ kernel functions. Each element in the buffer contains the $d$-dimensional kernel center and a weight associated with the kernel. In some cases, the algorithm is also allowed to choose kernel bandwidth. When a data element arrives from the stream, we \textit{merge} the element with one of the kernels in the buffer and update the center and weight~\cite{zhou2003m}. 
The $M$ kernel centers can be found via online clustering~\cite{heinz2008cluster}, self-organizing maps~\cite{cao2012somke}, and a variety of similar methods for the Gaussian, Epanechnikov and Laplace kernels. The original $M$-kernel approach only supports 1-dimensional data~\cite{zhou2003m}, but there are extensions to higher dimensions~\cite{xu2014dm}. The update and memory cost is $O(Md)$ for all merging algorithms. Unfortunately, such methods do not scale well since roughly one kernel center is needed for each cluster in the dataset~\cite{heinz2008cluster}. Unless the dataset lies in a very low dimensional subspace, $M$ grows very quickly with $d$. Furthermore, kernel merging requires us to store dense $d$-dimensional vectors as part of the sketch. 


\paragraph{Sampling} Another family of algorithms is based on a sampling approach. Here, we begin by sampling a set of points from the dataset. Then, we assign a weight to each of the samples. Finally, we approximate the full KDE by computing the weighted KDE over the set of samples. Sampling methods are attractive in practice because they retain the data summarizing capability of merging methods without needing to store dense high dimensional centroids. Sampling is also faster than merging because we can directly select the kernel centers from the dataset and do not need online clustering. Sampling algorithms differ mainly by the process used to find the samples and the weights. The simplest example is random sampling, where we uniformly select samples from the dataset and assign the same weight to each sample. Random sampling is highly efficient and can operate in the streaming environment, but it requires a relatively large number of samples for an accurate estimate. 

The primary research focus in this area has been to find adaptive sampling procedures that require fewer samples to approximate the KDE. By reducing the number of kernel evaluations, we improve both the query cost and the sketch size. Since adaptive sampling is often expensive, most methods generate a sketch that will work for any query. Sampling procedures based on herding~\cite{chen2012super} k-centers~\cite{cortes2016sparse}, and a variety of other cluster-based approaches are available in the literature. These methods operate offline since efficient adaptive sampling on streaming data is a challenging problem. Recently, locality-sensitive hashing has been used as a fast adaptive sampler for the KDE problem~\cite{spring2017new,charikar2017hashing}. In particular, the hashing-based estimator (HBE) introduced by~\cite{charikar2017hashing} has strong theoretical guarantees for KDE, even in high dimensions. Recent experiments have rigorously validated the HBE framework in practice~\cite{siminelakis2019rehashing}. This method has also been extended to work with pairwise summations over a much larger class of convex functions~\cite{charikar2018multi} but has not yet been extended to work on streaming data.

\paragraph{Coresets} A coreset of a dataset is a small collection of sampled points that can be used to approximate functions over the dataset. Coresets are attractive for many applications because they come with strong theoretical error guarantees and often have a deterministic construction process. Coresets for kernel density estimation exist, but have $O(N^2)$ pre-processing time~\cite{phillips2018near}. There has been some progress toward efficient coresets on streaming data~\cite{karnin2019discrepancy}, but the memory requirement for KDE remains at least $O(\frac{d}{\epsilon^2})$. Note that the $\epsilon$ dependency can be reduced in low dimensions. 

\section{Background}
In this section we introduce preliminary material on locality sensitive hash functions and sketching algorithms. 
\subsection{Locality Sensitive Hashing}
A hash function $h(x) \mapsto \{1,...,R\} \in \mathbb{Z}$  is a function that maps an input $x$ to an integer in the range $[1,R]$. A locality sensitive hash (LSH) family is a family of hash functions with the following property: Under the hash mapping, similar points have a high probability of having the same hash value. The formal definition of a LSH family is as follows~\cite{indyk1998approximate}. 
\begin{definition} $(R, cR, k_1, k_2)$-sensitive hash family \\ 
A family $\mathcal{H}$ is called $(R, cR, k_1, k_2)$-sensitive with respect to a distance function $d(\cdot,\cdot)$ if the following properties hold for $h \in \mathcal{H}$ and any two points $x$ and $y$: 
\begin{itemize}
    \item If $d(x,y) \leq R$ then $\text{Pr}_{\mathcal{H}}[h(x) = h(y)] \geq k_1$
    \item If $d(x,y) \geq cR$ then $\text{Pr}_{\mathcal{H}}[h(x) = h(y)] \leq k_2$
\end{itemize}
\end{definition}
The two points $x$ and $y$ are said to \textit{collide} if $h(x) = h(y)$. We will use the notation $k(x,y)$ to refer to the collision probability $\text{Pr}_{\mathcal{H}}[h(x) = h(y)]$. The collision probability is bounded and symmetric: $0\leq k(x,y)\leq 1$, $k(x,y) = k(y,x)$, and $k(x,x) = 1$. It should be noted that for any positive integer $p$, if there exists an LSH function $h(\cdot)$ with $k(x,y)$, then the same hash function can be independently concatenated $p$ times to obtain a new hash function with collision probability $k^p(x,y)$. If $h(\cdot)$ had the range $[1,R]$, then the new LSH function has the range $[1,R^p]$. We will also use the rehashing trick. If a LSH function $h(\cdot)$ has collision probability $k(x,y)$ and we hash the LSH values to a finite range $[1,R]$ using a universal hash function, then the new hash function is locality sensitive with collision probability $k(x,y)\frac{R - 1}{R} + \frac{1}{R}$. The rehashing trick works even for LSH functions that originally had an infinite range.

\subsection{LSH Kernels}
For many LSH functions, the collision probability forms a positive semi-definite radial kernel. We will refer to such kernels as \textit{LSH kernels}. To ensure the existence of a LSH kernel for the LSH functions that we consider in this paper, we will require an additional LSH property that is slightly stronger than the one given by~\cite{indyk1998approximate}. We assume that $k(x,y) \propto f(d(x,y))$, where $f(\cdot)$ is monotone decreasing. This condition is sufficient for $k(x,y)$ to be a positive semi-definite kernel. Most of the LSH functions in the literature satisfy the monotone decreasing collision property. Examples include MinHash, signed random projections, bit sampling and the popular $p$-stable LSH scheme for the Euclidean distance. As a result, there are many LSH kernels available for practical applications. Although the closed-form analytic expressions are often complicated, most LSH kernels look and behave similarly to well-known kernels.

\subsection{Repeated Array-of-Counts Estimator (RACE)}
While LSH was originally introduced for the high dimensional nearest-neighbor search problem, the technique has also recently been applied to unbiased statistical estimation via adaptive sampling for a variety of functions~\cite{spring2017new,charikar2017hashing}. Our KDE method will use the RACE algorithm, which views LSH as a slightly different kind of statistical estimator~\cite{luo2018arrays}. The RACE algorithm compresses a dataset $\mathcal{D}$ into an array $A$ of $L \times R$ integer counters, where each of the $L$ rows of $A$ is an ACE data structure~\cite{luo2018arrays}. To add an element $x\in \mathcal{D}$ to $A$, we hash $x$ using a set of $L$ independent LSH functions $\{h_1(x), ..., h_L(x)\}$. Then, we increment the counter $A[i,h_i(x)]$ for $i = \{1,...,L\}$. Thus, each counter records the number of elements that hased to the corresponding LSH bucket. The main result of~\cite{luo2018arrays} is as follows. 
\begin{theorem} ACE Estimator \\ 
\label{thm:RaceEstimator}
Given a dataset $\mathcal{D}$, a LSH family $\mathcal{H}$ with finite range $[1,R]$ and a parameter $p$, construct a LSH function $h(x) \mapsto [1,R^p]$ by concatenating $p$ independent hashes from $\mathcal{H}$. Let $A$ be an ACE array constructed using $h(x)$. Then for any query $q$, 

$$ \E[A[h(q)]] = \sum_{x \in \mathcal{D}} k^p(x,q)$$

\end{theorem}
The key insight for KDE is that ACE is an unbiased estimator for LSH kernels. A tight variance bound for this estimator was proved by~\cite{coleman2019race}.
\begin{theorem} ACE Estimator Variance \\ 
\label{thm:RaceEstimatorVariance}
Given a query $q$, the variance of the ACE estimator $A[h(q)]$ obeys the following inequality
$$\var( A[h(q)] ) \leq \left(\sum_{x \in \mathcal{D}} k^{\frac{p}{2}}(x,q)\right)^2$$
\end{theorem}
These results suggest that repeated ACEs, or RACEs, can estimate the KDE with very low relative error given sufficiently large number of repetitions. The difference between our work and the work contained in~\cite{coleman2019race} and~\cite{luo2018arrays} is that we consider a much wider class of LSH functions than those considered in either paper. Critically, both papers considered only \textit{finite-range} LSH functions. In this paper, we will show that this is a special and easy case for kernel density estimation with RACE. 

\section{Kernel Density Estimation with RACE}
We will now describe our proposed kernel density algorithm. Our algorithm extends the ACE method from ~\cite{luo2018arrays} to work for the approximate KDE problem. We focus on the problem of obtaining an accurate estimate of the pairwise kernel summation rather than anomaly classification on data streams. Our algorithm is designed to handle arbitrary LSH functions, while the original ACE method only used signed random projections. We also introduce some algorithmic changes. For instance, we use a median-of-means procedure rather than an average to bound the failure probability of the randomized query algorithm. We also prove variance bounds to limit the number of repetitions needed for an accurate estimate. Otherwise, our method inherits all of the desirable practical properties from ACE.

\begin{figure*}[t]
\vspace{-0.15in}
\mbox{
\centering
\hspace{-0.2in}
\includegraphics[width=6in,keepaspectratio]{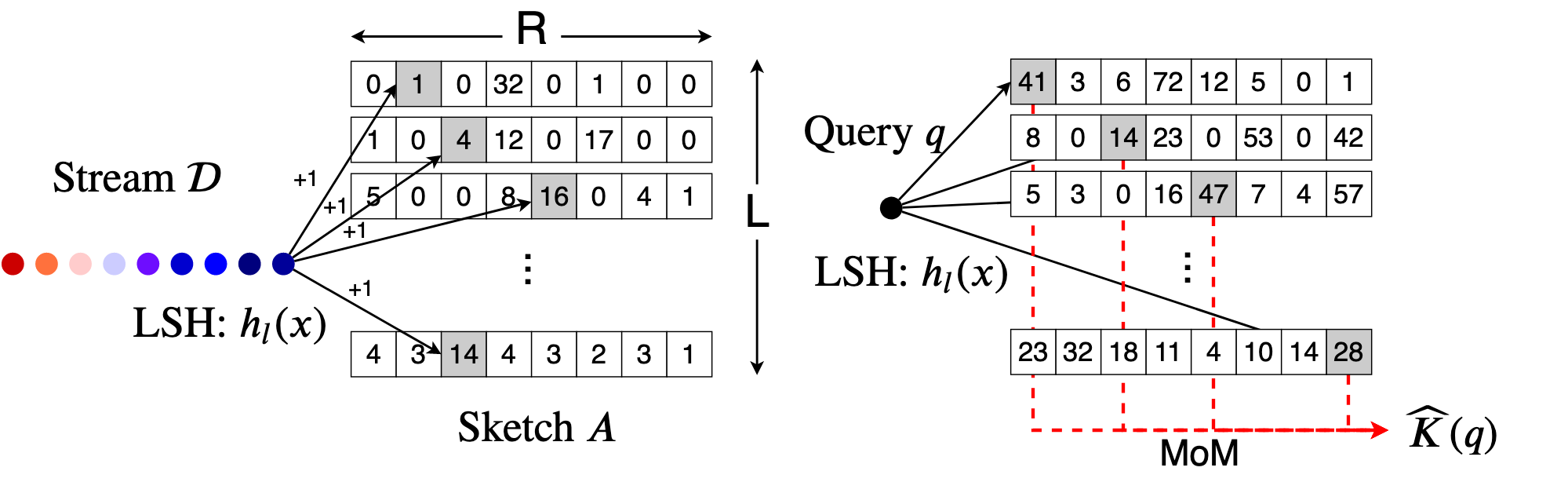}
}
\vspace{-0.15in}
\caption{ Illustration of the pre-processing / sketching (left) and querying (right) parts of Algorithm~\ref{alg:sketch}. While sketching the data stream $\mathcal{D}$, we hash each element with $L$ LSH functions and we increment the corresponding array locations. To query the sketch, we hash the query with the same LSH functions and find the mean of the count values at those locations.} 
\label{fig:raceAlgo}
\end{figure*}

\begin{algorithm}[t]
\begin{algorithmic}
\STATE {\bf Online pre-processing:}
\STATE {\bf Input:} Dataset $\mathcal{D}$, LSH kernel parameters $\alpha, p$, ACE repetitions $L$ and hash range $R$
\STATE {\bf Output:} RACE sketch $A \in \mathbb{Z}^{L\times R}$
\STATE {\bf Initialize:} $L$ independent LSH functions $\{h_1, ..., h_L\}$ with bandwidth $\alpha$, power $p$ and range $R$
\STATE $A \leftarrow \mathbf{0}^{L\times R}$
\FOR{$x \in \mathcal{D}$}
    \FOR {$l$ in $1$ to $L$}
        \STATE Increment $A[l,h_l(x)]$
    \ENDFOR
\ENDFOR
\STATE
\hrulefill
\STATE {\bf Online query:}
\STATE {\bf Input:} Query $q$, RACE sketch $A$, LSH kernel parameters $\alpha, p$, ACE repetitions $L$, hash range $R$, number of means $N_M$ 
\STATE {\bf Output:} Approximate kernel density $\widehat{K}(q)$
\STATE {\bf Initialize:} $L$ independent LSH functions $\{h_1, ..., h_L\}$ with bandwidth $\alpha$, power $p$ and range $R$ using the same random seed as in pre-processing
\STATE {\bf Initialize:} Vector of means $M\leftarrow \mathbf{0}^{N_M}$
\STATE $\widehat{K}(q)$
\FOR {$l$ in $1$ to $L / N_M$}
    \FOR {$m$ in $1$ to $N_M$}
        \STATE $M[l] = M[l] + \frac{1}{N_M} A[l,h_l(q)]$  
    \ENDFOR
\ENDFOR
\STATE $\widehat{K}(q) \leftarrow \mathrm{median}(M)$
 \end{algorithmic}
  \caption{Online KDE sketch construction and querying algorithm}
 \label{alg:sketch}
\end{algorithm}

\paragraph{Intuition} The KDE of a query is roughly a measure of the number of nearby elements in the dataset. Each row of the ACE array is a noisy representation of this information. Nearby elements are likely to collide with each other and with the query, while collisions with faraway elements are unlikely. Therefore, each cell in the ACE array can be thought of as a noisy or soft data cluster with a weight equal to the count value. By averaging over several ACEs, we reduce the effect of the noise. Since the ACE estimates sharply concentrate around the mean, only a few repetitions are needed to provide a good KDE approximation.

The proposed algorithm has several desirable features in practice. First, the $L$ hash function computations can be done in parallel during the online sketching and querying procedures. Our sketch is simple to understand and implement, as $A$ consists of an array of integer counters. Although we do not do this in our evaluation, additional integer-specific memory optimizations can be applied to $A$. For example, $A$ often consists of many entries that are either zero or approximately the same (large) value. We can save memory by storing $A$ in sparse format and by packing $A$ into a compressed short array using delta coding. 

Furthermore, our sketch is mergeable without increasing the error. Suppose we use Algorithm~\ref{alg:sketch} to create $A_1$ from $\mathcal{D}_1$ and $A_2$ from $\mathcal{D}_2$ using the same LSH functions. To construct an estimator for $\mathcal{D}_1 \cup \mathcal{D}_2$, we only need to add the counters from $A_1$ and $A_2$. This property is critical for large-scale distributed systems because $A_1$ and $A_2$ can be constructed on separate systems. As long as the two RACEs are construced using the same LSH function seed and RACE parameters, they can be merged or updated dynamically at any time. 

\section{Theory}
Here we present a theoretical analysis of our algorithm. We start with a simple case in Section~\ref{sec:AngularRACE}. We extend this argument for all LSH kernels in Section~\ref{sec:ApproxLSHKernels} and finally show how to estimate arbitrary kernels in Section~\ref{sec:ApproxAllKernels}. 

\subsection{RACE for the Angular Kernel}
\label{sec:AngularRACE}
Suppose we wish to estimate the LSH kernel sum 
\begin{equation}
\label{eqtn:KDE}
K(q) = \frac{1}{|\mathcal{D}|}\sum_{x \in \mathcal{D}} k^p(x,q)
\end{equation}
For concreteness, consider the signed random projection (SRP) LSH family, a set of functions $h_w(\mathbf{x}): \mathbb{R}^n \mapsto \{-1,1\}$.
$$h_w(\mathbf{x}) = \mathrm{sign}(\mathbf{w}^{\top}\mathbf{x})$$
The collision probability of $\mathbf{x}$ and $\mathbf{y}$ under SRP is 
$$ k(\mathbf{x},\mathbf{y}) = 1 - \frac{1}{\pi}\theta(\mathbf{x},\mathbf{y})$$
where $\theta(\mathbf{x},\mathbf{y})$ is the angle between $\mathbf{x}$ and $\mathbf{y}$. 
It should be noted that this collision probability is equivalent to the angular kernel given in~\cite{choromanski2017unreasonable}. Angular kernel methods have applications in information retrieval, anomaly detection~\cite{luo2018arrays}, natural language processing, image analysis and spectral processing~\cite{honeine2010angular}. We can directly apply the median-of-means procedure to Theorem~\ref{thm:RaceEstimatorVariance} to get an upper bound on the number of ACE repetitions $L$ needed for a $1 \pm \epsilon$ multiplicative approximation of $\KDE(\mathbf{q})$. Since the range of the SRP family is finite ($R = 2$), each repetition only requires $2^p$ memory. Generalizing this result to apply to all LSH functions with finite range $R$, we obtain the following corollary to Theorem~\ref{thm:RaceEstimatorVariance}. 

\begin{corollary} RACE Memory Bound\\ 
\label{cor:RaceMemoryBound}
Suppose we are given a LSH function with range $R$. Then a RACE array with 
$$ L = O\left(\left(\frac{\widetilde{K}(q)}{K(q)}\right)^2\frac{1}{\epsilon^2} \log \frac{1}{\delta}\right)$$
independent rows provides a $1 \pm \epsilon$ multiplicative approximation to $K(q)$ with probability $1 - \delta$ in space $O(LR^p)$, where $\widetilde{K}(q)$ is
$$ \widetilde{K}(q) = \frac{1}{|\mathcal{D}|}\sum_{x \in \mathcal{D}} k^\frac{p}{2}(x, q) $$
\end{corollary}
\subsection{Approximation of LSH Kernels}
\label{sec:ApproxLSHKernels}
Corollary~\ref{cor:RaceMemoryBound} is only meaningful when the range $R$ of each LSH function is bounded. However, many useful LSH kernels do not have this property. For example, the $p$-stable LSH functions for the Euclidean and Manhattan distance (or L2 and L1 LSH, respectively) have infinite range, so the corresponding LSH kernels are more difficult to estimate. These functions are defined in~\cite{datar2004locality}. The L1 LSH function has 
$$ k(\mathbf{x},\mathbf{y}) = \frac{1}{\pi} \mathrm{atan}\left(\frac{\sigma}{||\mathbf{x}-\mathbf{y}||_1}\right) - \frac{||\mathbf{x}-\mathbf{y}||_1}{\pi \sigma} \ln \left(1 + \left(\frac{\sigma}{||\mathbf{x}-\mathbf{y}||_1}\right)^2\right)$$
and the L2 LSH function has 
$$ k(\mathbf{x},\mathbf{y}) = -\frac{1}{2} \mathrm{erf}\left(\frac{-\sigma}{||\mathbf{x}-\mathbf{y}||_2\sqrt{2}}\right) - 2\frac{||\mathbf{x}-\mathbf{y}||_2}{\sigma \sqrt{2\pi}}\left(1 -\exp\left(\frac{-\sigma^2}{2||\mathbf{x} - \mathbf{y}||_2^2}\right)\right)$$
While the closed-form expression for $k(\mathbf{x},\mathbf{y})$ may be complicated, these two functions are nearly identical to the exponential kernel in $d(\mathbf{x},\mathbf{y})$. As a result, they have an unbounded support. To limit the memory needed by each ACE array, we use the rehashing trick to restrict the range of each array to exactly $R$ counts. Under these conditions, RACE will estimate pairwise sums of $k^p(x,q)\frac{R - 1}{R} + \frac{1}{R}$. The new collision probability is still a positive definite kernel which can be estimated via Corollary~\ref{cor:RaceMemoryBound}. However, suppose we want to estimate the original KDE (Equation~\ref{eqtn:KDE}). The estimator presented by ~\cite{coleman2019race} actually estimates the quantity
$$\frac{|\mathcal{D}|}{R} \sum_{x\in\mathcal{D}} k^p(x,q)\frac{R - 1}{R}$$
This is is now a biased estimator, so we require a different estimator. 

\begin{theorem} Rehashed RACE Estimator \\
\label{thm:RehashedRaceEstimator}
Let $A$ be a rehashed ACE data structure. That is, suppose $A$ is constructed using a LSH function that has been rehashed to the range $[0,R]$. Let
$$ \widehat{K}(q) = \left(\frac{A[h(q)]}{|\mathcal{D}|} - \frac{1}{R}\right)\frac{R}{R-1}$$
Then $\E[\widehat{K}(q)] = \mathrm{KDE}(q)$ and
$$ \var(\widehat{K}(q)) \leq \left(\frac{R}{R-1}\right)^2\left( \sqrt{\frac{R-1}{R}} \widetilde{K}(q) + \frac{1}{\sqrt{R}}\right)^2 $$
\end{theorem}
Using Theorem~\ref{thm:RehashedRaceEstimator} and the same median-of-means procedure as before, we can bound the number of ACE repetitions needed for an accurate KDE approximation. 
\begin{corollary} Rehashed RACE Memory Bound\\ 
\label{cor:RehashedRaceMemoryBound}
Suppose we rehash a LSH function with collision probability to a finite range $R$. Then a RACE array with 
$$ L = O\left(\left(\frac{R}{R-1}\right)^2\frac{1}{K(q)^2} \frac{1}{\epsilon^2}\log \frac{1}{\delta}\right)$$
independent rows provides a $1 \pm \epsilon$ multiplicative approximation to $K(q)$ with probability $1 - \delta$ in space $O(LR)$. 
\end{corollary}

A comparison of Corollary~\ref{cor:RehashedRaceMemoryBound} with Corollary~\ref{cor:RaceMemoryBound} shows that rehashing has a price. Since $\widetilde{K}$ and $K$ are of approximately the same order of magnitude, $\frac{1}{K^2}>\left(\frac{\widetilde{K}}{K}\right)^2$. Therefore, we need more ACE repetitions when we use rehashing to limit the space required by each array. Equivalently, we lose precision when estimating low values of $K(q)$ with a rehashed RACE structure. The key insight is that LSH kernels are fundamentally easier to approximate when the underlying LSH function has finite range. This corresponds to the support of the kernel. If a LSH kernel has infinite support, then it has infinite range and is harder to estimate. On the other hand, if a LSH function has a bounded range as with random projections or bit sampling LSH, the corresponding LSH kernel will have finite support. 

\subsection{Approximation of Arbitrary Kernels}
\label{sec:ApproxAllKernels}
We have shown that RACE approximates the KDE for the special class of LSH kernels, but our method also applies to all radial kernels of the form $k(d(x,q))$, where $d(x,q)$ is a distance function, albeit without formal error bounds. 

\begin{theorem} Arbitrary Kernels (Informal)\\
A small collection of RACE sketches can estimate kernels $k(x,q)$ of the form $k(d(x,q))$. 
\end{theorem}
\begin{proof}
Let the LSH kernel collision probability be $f(d(x,q))$ and note that $f$ is monotone decreasing. Therefore, it has an inverse, and we can express $k(x,q)$ as $k(f^{-1}(f(d(x,q))))$. Let $z(x,q) = f(d(x,q))$ and note that RACE can estimate $Z = \sum_{x \in \mathcal{D}} z(x,q)^p$ for any power $p$. Apply the procedure from~\cite{itskov2012taylor} to compute the Taylor series coefficients of $g(z) = k(f^{-1}(z))$ and construct one RACE for each power of $z$. Output the linear combination of RACE estimates. 
\end{proof}

While one can construct an unbiased estimator for any radial kernel using the series expansion, it is difficult to bound the error of the estimator. Since the variance now depends on the size of each coefficient and the quality of the Taylor approximation, it is much more complicated to prove error bounds. Empirically, we found that this procedure generates reasonably good approximations to $d(x,q)$ from $f(x,q)$. For instance, we constructed a Taylor series inverse approximation with 10 coefficients about the point $d(x,q) = 1$ for the L1 LSH collision probability with $\sigma = 4$. Provided that $||\mathbf{x} - \mathbf{q}||_1 < 10$, we can estimate $d(x,q)$ with less than 10\% error using the approximation. With the correct coefficients, we can approximate arbitrary kernels. Although this idea is theoretically appealing, in practice it is much more effective to use regularized linear regression on the $k(d(x,q))$ curves to obtain a low-norm set of coefficients. Input a set of kernels that RACE can estimate and perform regression to approximate the desired kernel. 

\section{Discussion}

\subsection{Computation-Memory Tradeoff}
Corollary~\ref{cor:RehashedRaceMemoryBound} seems to indicate that we should choose the rehashing range to be very small so that the memory is minimized. From a memory perspective, the optimal range minimizes $L\times R$ subject to $R \geq 2$. If we solve this optimization problem, we find that the optimal $R$ for memory size is $R = 3$. However, there are practical trade-offs associated with choosing the rehashing range. Updating and querying a RACE sketch has time complexity $O(L)$, since $L$ hash functions need to be computed. To minimize the update and query time, we should choose the range to be as large as possible to minimize $L$. This results in a memory-computation trade-off. 

One may choose any two of the following three parameters: error, memory, and update cost. We can understand this trade-off with a simulated experiment. We first generate a clustered dataset with half a million points. We construct multiple RACE sketches with different sets of parameters. We find that a RACE with $R = 3$ and $L = 10\times 10^3$ had $0.5\%$ relative error. The same error is attainable by a RACE sketch with $R = 4\times 10^3$ and $L = 2\times 10^3$. The first sketch is 260x smaller than the second one, but each update requires 5x as many hash computations. From this small experiment, we see that extreme compression is possible but comes at an increased computational cost. 

Fortunately, the hash computations are not terribly expensive. In practice, $L$ can usually be smaller than 200 for good results ($\epsilon < 5\%$). The LSH computations are trivial to parallelize and can be implemented in hardware. Many LSH functions of interest simply require an inner product computation and are therefore well-suited to implementation on GPU. Furthermore, if we use sparse arrays to implement RACE, we effectively have an unlimited rehashing range without dramatically increasing the memory. Since many of the RACE counts are usually zero, this approach can decrease both the memory and the runtime. 

\subsection{Privacy}
RACE has unique privacy-preserving properties. Specifically, we do not store any elements or data attributes from the dataset. The entire dataset is streamed into the RACE sketch without ever being stored anywhere. The sample-based approaches described previously require data structures that explicitly store parts of the dataset. Kernel merging methods are more amenable to privacy because they store conglomerated summaries of the dataset, but we still store averages of user data attributes. RACE is constructed via a randomized hash function that does not need to store or merge data attributes at all. 

Although a complete privacy analysis of RACE is beyond the scope of this paper, there is evidence that differentially private ACE structures can be constructed using modified hash functions~\cite{luo2018arrays}. This technique applies to random projection hashes, including the SRP and $p$-stable hash functions discussed previously. By adding Gaussian noise to the random projection, we obtain a LSH code with differential privacy~\cite{kenthapadi2012privacy}. We expect that private RACE data structures will retain their utility for most applications, since the Johnson-Lindenstrauss privacy mechanism does not substantially perturb the LSH collision probability. 

\section{Experiments}
Our goal in this section is to explore the memory-accuracy tradeoff of RACE compared with baselines. We provide two sets of experiments. First, we use the $p$-stable Euclidean LSH kernel. While the analytic expression is complicated, this LSH kernel strongly resembles the well-known exponential kernel with an analogous bandwidth parameter. Next, we estimate the angular kernel density using RACE. 

\subsection{Experimental Setup}

\paragraph{Storage:} For our sampling baselines, we store samples in the smallest format possible. For datasets such as webspam and url, we store sparse arrays. For the dense audio and imaging datasets we report results for standard arrays. We store all integers and floating-point numbers in standard 32-bit types. Even though RACE counts can often fit into short integers, we do not perform aggressive memory optimization because we aim to provide a fair comparison that validates our theory. 

\paragraph{Rehashing:} We use rehashing to bound the range of our $p$-stable LSH functions. In practice, the majority of the RACE counts are zero even after rehashing. According to the trade-off discussed earlier, we can produce very small sketches with dense arrays by rehashing to a small range. However, to keep computation costs comparable with other methods we choose to use as large a rehashing range as possible. Therefore, we use the maximum unsigned integer range and implement RACE using sparse arrays. Note that this satisfies  Corollary~\ref{cor:RehashedRaceMemoryBound} with $R$ equal to $2^{32}$. For the angular kernel, rehashing and sparse storage are not necessary since $R = 2$. 

\paragraph{Bandwidth Selection:} The LSH kernel accepts a bandwidth parameter that can be tuned. Bandwidth selection is an important topic in high-dimensional statistical modeling, but the purpose of our experiment is to validate our theory and show that RACE sketches are practical. Our analysis predicts that RACE will outperform sample-based methods when $d > \frac{1}{K(q)}$, so we chose the bandwidth parameters that characterize the relationship for a wide range of $K(q)$. For our experiments with angular kernel density, we use the power $p = 1$. 

\paragraph{KDE Queries:} We use held-out test data rather than random queries in our KDE experiments. In high dimensions and with very sparse inputs, random queries tend to have extremely low kernel densities. Low-density queries are usually not meaningful in practice and do not provide an informative comparison of our baseline methods because of the $\KDE^{-1}$ factor for all methods in Table~\ref{table:relatedwork}. It is difficult for all methods to provide a multiplicative $1\pm\epsilon$ approximation when $\KDE$ is very small. Therefore, we use queries that come from the same distribution as the dataset. This allows us to characterize our method for a wider range of $K(q)$ and is a better representation of real performance.

\subsection{Baselines}
We implemented random sampling, sparse kernel approximation, and the hash-based sketch. We do not include kernel herding since we found the $O(|\mathcal{D}|^2)$ sketching cost to be prohibitive. We do not compare against the M-kernel or cluster kernel methods due to the prohibitive storage requirement. These methods are difficult to evaluate in high dimensions and require storing \textit{dense} high dimensional vectors. We briefly describe our baseline methods: 
\begin{enumerate}
	\item \textbf{Random Sampling (RS):} Using reservoir sampling, we select a set of points from the dataset and weight them equally. This is the only baseline that can operate in the streaming setting.
	\item \textbf{Hash-Based Sketch (HBS):} This is a two-pass method that samples points using LSH. First, we use a LSH function to hash each point in the dataset into a hash table. In the second pass, we sample points from the hash table and weight them according to the size of each hash bucket. This method accepts the number of tables and the LSH function bandwidth as hyperparameters. We use the settings recommended in the paper. 
	\item \textbf{Sparse Kernel Approximation (SKA):} Sparse kernel approximation samples points from the dataset using the greedy k-centers algorithm. We randomly choose the first point, then iteratively select points from the dataset that are far away from points in our sample. Each point in the sketch is weighted so that the KDE at that point is correct. 
\end{enumerate}

\subsection{Datasets}

\begin{table*}[t]
\begin{center}
\begin{tabular}{ l | l | l | l | l | l | l }
\hline
Dataset & $N$ & $d$ & $C$ & Kernel & $\sigma$ & Description \\
\hline
Intervalgrams & 92k & 5000 & 2400 & L2 & 0.05 & Audio spectrogram features \\
Audio Boxes & 97k & 5000 & 3174 & L2 & 0.016 & Audio signal codebook features \\
Webspam & 350k & 2.3M & 3729 & L2 & 0.02 & Website integrity features \\
URL & 2.3M & 3.2M & 115 & L2 & 15 & URL reputation features \\
PA & 783k & 102 & 102 & Angular & -- & ROSIS sensor data for Pavia, Italy \\
KSC & 314k & 176 & 176 & Angular & -- & AVIRIS data for Kennedy Space Center \\
IP & 21k & 220 & 220 & Angular & -- & AVIRIS data for Indian Pines, IN \\
SA & 111k & 224 & 224 & Angular & -- & AVIRIS data for Salinas, CA \\
\hline
\end{tabular}
\end{center}
\caption{Dataset information. Each dataset has $N$ vectors with $d$ dimensions. Since $d$ is not the true cost to store a sample, we also report the sparsity of each vector as the average number of nonzeros. If this number is smaller than $d/2$, we store samples in sparse format. Here, we use L2 to refer to the $p$-stable LSH kernel for the Euclidean distance. For all experiments, $p = 1$.}
\label{table:datasets}
\end{table*}

We chose to evaluate our method on datasets from real-world application domains that are good candidates for compression using sampling or RACE sketches. Table~\ref{table:datasets} contains information about our datasets and experiments. For the $p$-stable LSH kernel, we target applications in large-scale web analysis and audio processing. We evaluate RACE on large, high-dimensional and sparse features from the Webb spam corpus~\cite{wang2012evolutionary} and URL reputation datasets~\cite{ma2009identifying}. We also use a set of pre-processed audio features from a large-scale analysis of YouTube video data~\cite{madaniEtAl2013MLJ}. We use 10,000 held-out test points as queries.  

For the angular kernel, we use hyperspectral image data where each image pixel corresponds to a frequency spectrum. The angular distance between two spectra is a meaningful class indicator for ground cover type and other important scene characteristics~\cite{honeine2010angular}. We obtained scenes from the AVIRIS and ROSIS sensors for our evaluation~\cite{PURR1947}. We treat each pixel as a high-dimensional data vector and apply the angular kernel using the same procedures as~\cite{honeine2010angular}. For our imaging datasets, we use 2,000 held-out test points as queries. 

\begin{figure}[t]
    \vspace{-0.15in}
    \centering
    \includegraphics[width=3in,keepaspectratio]{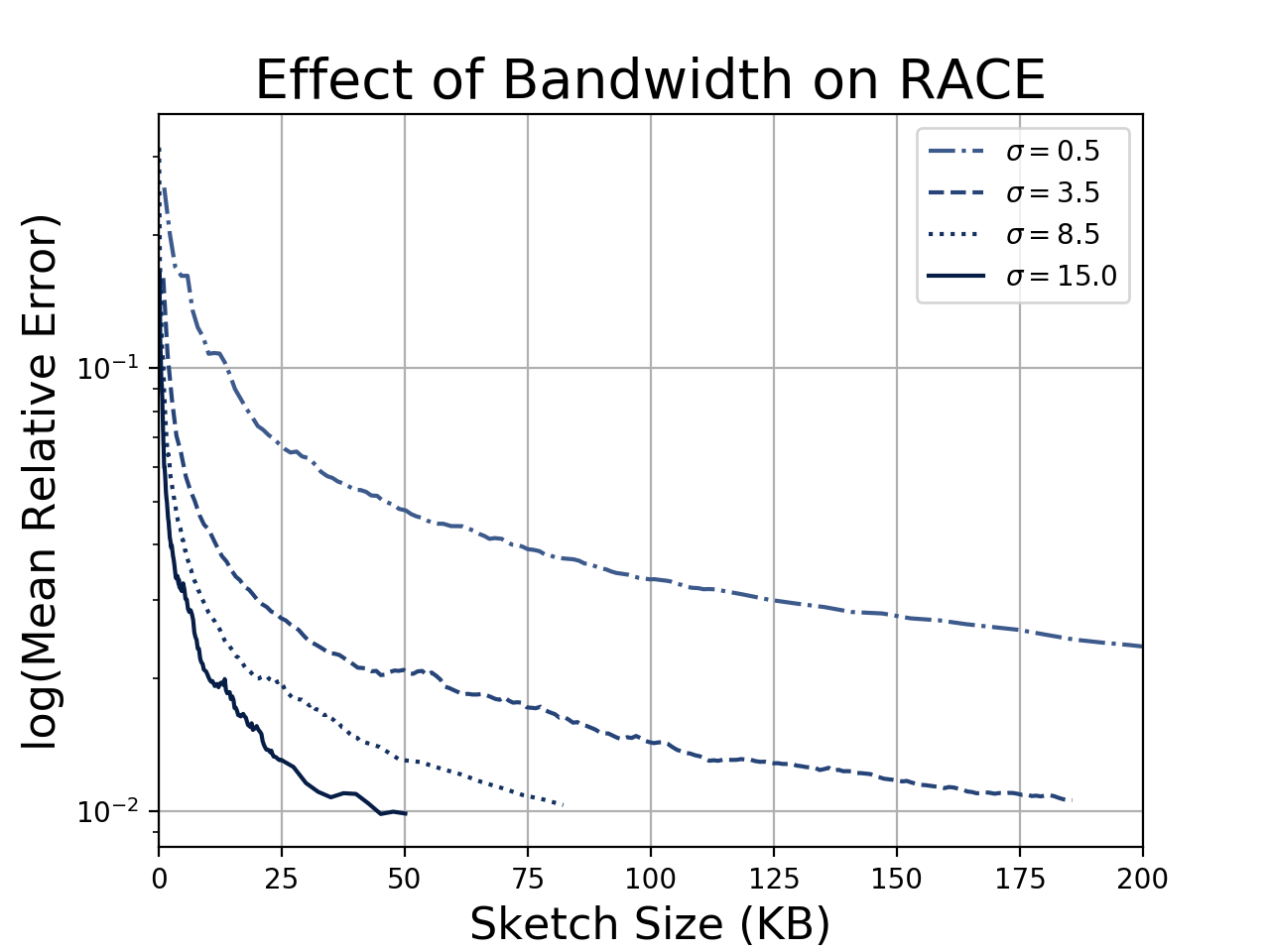}
    \caption{Plot of error vs sketch size for the URL dataset with varying $\sigma$. Increasing the bandwidth decreases the size required by the RACE sketch.}
    \label{fig:varyBW}
\end{figure}

\begin{figure}[t]
\vspace{-0.15in}
\mbox{
\centering
\hspace{-0.2in}
\includegraphics[width=3in,keepaspectratio]{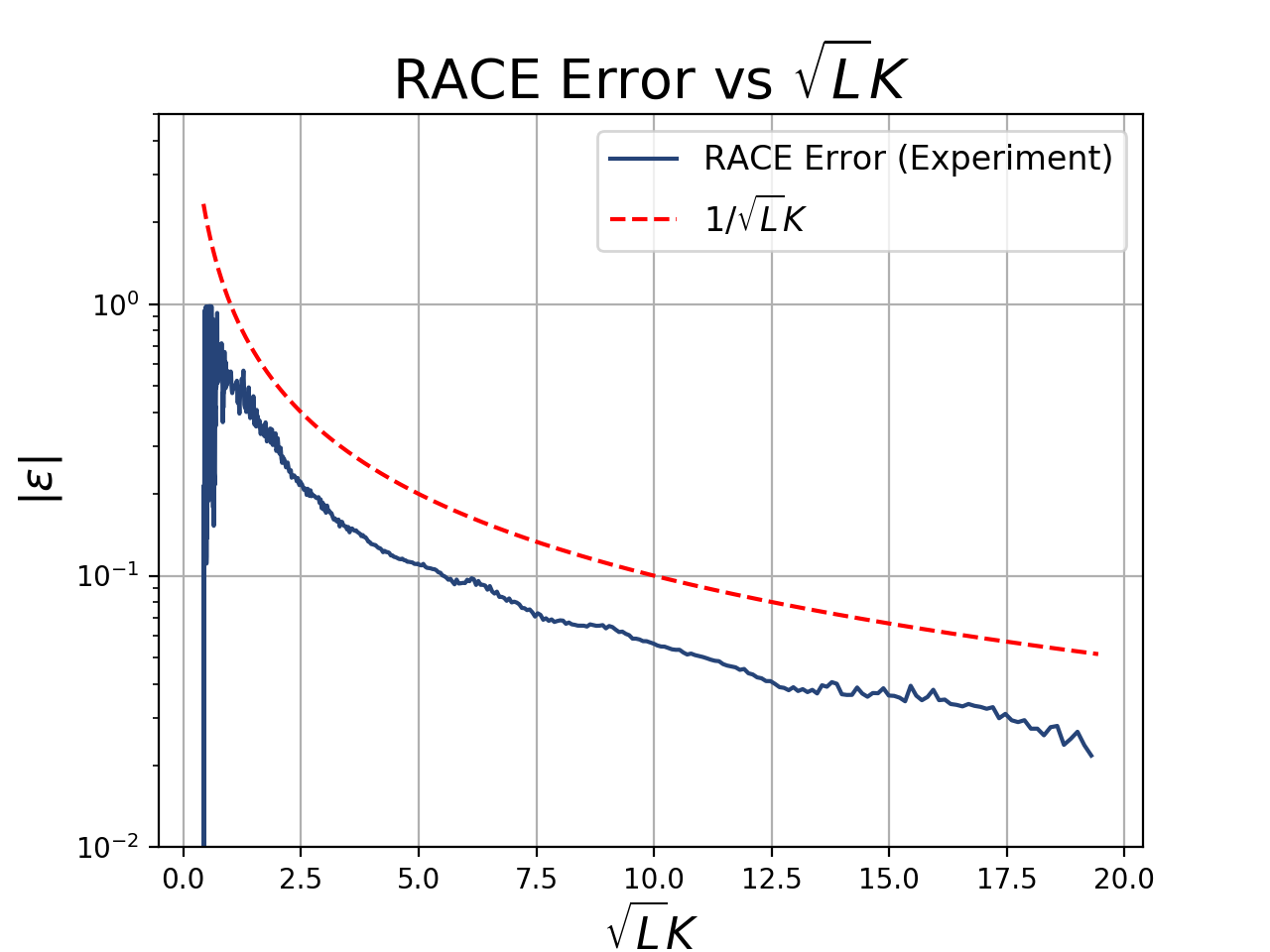}
}
\vspace{-0.15in}
\caption{Theoretical and empirical error measurements for the failure probability $\delta = 1\%$. Our theoretical results suggest that $|\epsilon| = O(1 / \sqrt{L} K(q) )$. This plot was obtained by binning the error for all URL dataset queries and sketches and finding the 99\textsuperscript{th} percentile of $|\epsilon|$ in each bin.} 
\label{fig:errorRate}
\end{figure}
\begin{figure*}[t]
\vspace{-0.15in}
\mbox{
\centering
\hspace{-0.2in}
\includegraphics[width=5.5in,keepaspectratio]{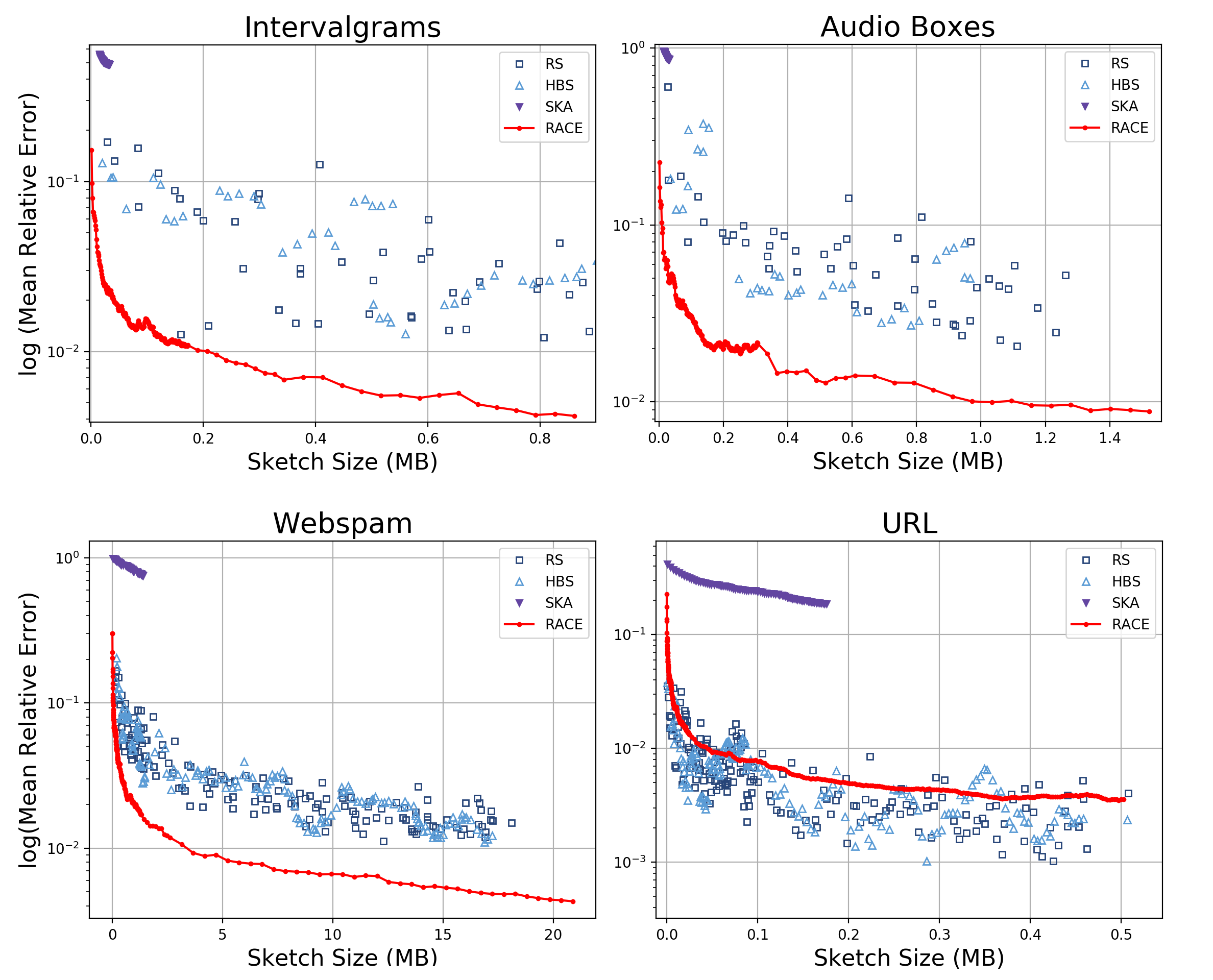}
}
\vspace{-0.15in}
\caption{ $(1 \pm \epsilon)$ multiplicative error using the $p$-stable LSH kernel with the datasets in Table~\ref{table:datasets}. Lower is better. The (mean, standard deviation) values of $K(q)$ were as follows. URL: (0.56, 0.03); Webspam: (0.0079, 0.0015); Intervalgrams: (0.57, 0.11); Audio Boxes: (0.19, 0.050). }
\label{fig:p_stable_sketches}
\end{figure*}

\begin{figure*}[t]
\vspace{-0.15in}
\mbox{
\centering
\hspace{-0.2in}
\includegraphics[width=5.5in,keepaspectratio]{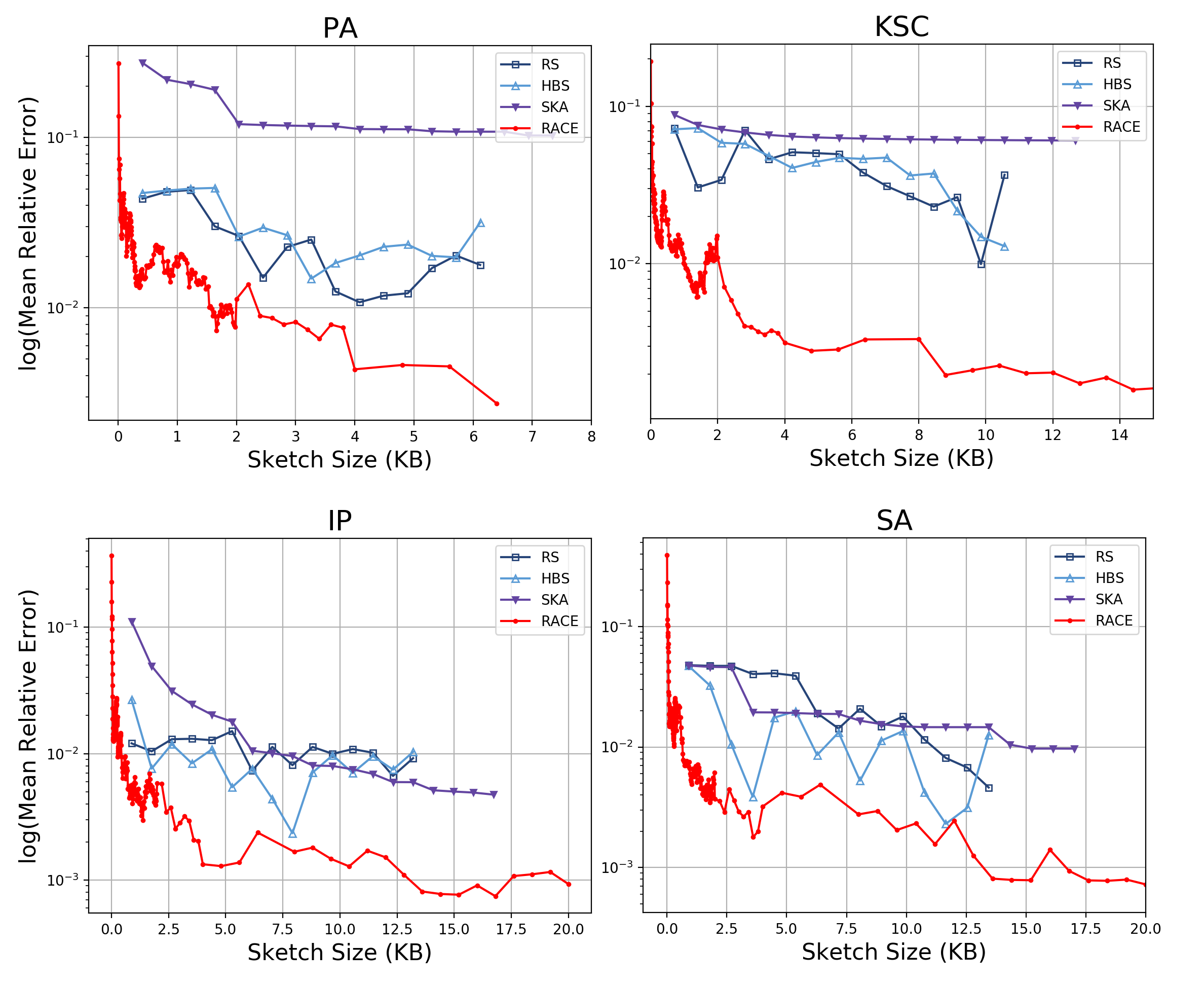}
}
\vspace{-0.15in}
\caption{ $(1 \pm \epsilon)$ multiplicative error using the $p$-stable LSH kernel (top) and Angular kernel (bottom) with the datasets in Table~\ref{table:datasets}. Lower is better. We used $p = 1$ for these experiments. Higher values of $p$ reduce the kernel densities and increase the range of the LSH function. The (mean, standard deviation) values of $K(q)$ were as follows. PA: (0.85, 0.05); KSC: (0.82, 0.09); IP: (0.95, 0.01); SA: (0.92, 0.02).} 
\label{fig:angular_sketches}
\end{figure*}

\subsection{Results}
Figure~\ref{fig:p_stable_sketches} and Figure~\ref{fig:angular_sketches} shows the results of our experiments. RACE consistently outperforms sample-based methods by factors of up to 10x. The URL dataset is the only exception. Corollary~\ref{cor:RehashedRaceMemoryBound} shows that RACE performs best when the sample storage cost $C \geq O(1 / K(q))$. Therefore, it was initially surprising that RACE performed poorly on this dataset, since $K(q)$ was relatively large for URL. However there are only 115 nonzeros per sample in the URL dataset. Therefore, sample-based methods on URL are relatively inexpensive in terms of memory when compared to RACE. Interestingly, RACE outperformed the baselines on the webspam dataset, where we used very low-density queries. It seems that given a high enough sample storage cost $C$, RACE will eventually beat sample-based methods in practice. The threshold seems to be approximately a few thousand dimensions for rehashed RACE estimators. 

Our experiments with the angular kernel show that LSH kernels corresponding to finite range LSH functions can be sketched using much less space. RACE outperformed the baselines even for the PA sensor data, which has only 102 dimensions. It should be noted that all of these datasets had relatively large values of $K(q)$, but the sample storage cost was also particularly low. Given the similar kernel density values and costs between the URL and hyperspectral imaging datasets, it is safe to conclude that finite-range LSH functions truly are easier to estimate. Our results also validate the accuracy of the anomaly detection experiments in~\cite{luo2018arrays} because we show that very few ACE repetitions are needed to get good density estimates with signed random projections.

We tried to use SKA with a variety of sketch sizes for our audio and web datasets. However, the greedy k-center sketching algorithm tends to select the sparsest elements from the dataset, leading to very small sketches. SKA was ultimately only competitive for our smallest two datasets, which is unsurprising given the theoretical results in Table~\ref{table:relatedwork}. Specifically, SKA is the only method that scales with $N$. Since the memory for SKA depends linearly on $K(q)$, we see improved performance for the URL and hyperspectral imaging datasets where $K(q)$ was large. RS and HBS were much more competitive and had similar performance in our tests. We observed that HBS performed slightly better for queries in sparse regions of the dataset. All of our baseline methods, including RACE, eventually yielded compression ratios of over 1000x when compared against the original dataset sizes. Most of the datasets in Figure~\ref{fig:p_stable_sketches} have uncompressed sizes that are over 1 GB. The imaging datasets are consistently larger than 1 MB. 

We also wanted to investigate how the bandwidth of the LSH kernel affects sketch performance under a fixed memory budget. Therefore, we created RACE sketches for a variety of $\sigma$ values on the URL dataset. The results are shown in Figure~\ref{fig:varyBW}. As expected, we require a larger sketch to represent the fine structure of low-bandwidth KDEs. Intuitively, increasing the bandwidth decreases the number of nonzero RACE counters because each LSH bin becomes wider. 

Finally, we validated our theoretical error guarantees with our experiments. Corollary~\ref{cor:RehashedRaceMemoryBound} suggests that given $L$ ACE repetitions, our estimate of $K$ has relative error $\epsilon$ where $|\epsilon| \leq O(1 / \sqrt{L}K)$ with high probability. We found the error for each query in the URL dataset and plotted the errors in Figure~\ref{fig:errorRate}. In practice, the 99\textsuperscript{th} percentile error has the asymptotic behavior predicted by our theoretical results and is bounded by the expression from Corollary~\ref{cor:RehashedRaceMemoryBound}. 

\section{Conclusion}
Data sketches for kernel density estimation currently require memory that is linear in the sample storage cost $C$. We leverage the recently-proposed RACE streaming algorithm to compress a dataset into an efficient sketch consisting of integer counters. This sketch can estimate the kernel density for a large and useful class of kernels. When $C \geq 1 / K$, RACE outperforms sample-based methods. In practice, RACE obtains compression ratios that are up to 10x better than sampling baselines. We believe that the simplicity and size of our RACE sketches will make them useful for data analysis in memory-constrained settings. 
\bibliographystyle{ACM-Reference-Format}
\bibliography{main}

\end{document}